\documentclass[useAMS,usenatbib]{mn2e}
\usepackage{graphicx}
\usepackage{amsmath}
\usepackage{makecell}
\usepackage{macros}
\usepackage{cancel}
\usepackage[usenames,dvipsnames,svgnames,table]{xcolor}
\usepackage{hyperref}
\definecolor{darkblue}{rgb}{0.0,0.0,0.3}
\hypersetup{colorlinks,breaklinks,
            linkcolor=darkblue,urlcolor=darkblue,
            anchorcolor=darkblue,citecolor=darkblue}
\usepackage{txfonts}
\usepackage[normalem]{ulem}
\DeclareSymbolFont{cmletters}{OML}{cmm}{m}{it}
\DeclareMathSymbol{v}{\mathalpha}{cmletters}{"76}

\newcommand{\Heavi}{\operatorname{H}}

\newcommand{\RedeclareMathOperator}[2]{\renewcommand{#1}{}\let#1\relax\DeclareMathOperator{#1}{#2}}

\newcommand\simless\lesssim
\newcommand\simgreat\gtrsim

\title[Bardeen-Petterson Alignment, Jets and Magnetic Truncation]{Bardeen-Petterson Alignment, Jets and Magnetic Truncation in GRMHD Simulations of Tilted Thin Accretion Discs}
\author[M. Liska et al.]{M. Liska$^1$\thanks{E-mail: matthewliska92@gmail.com}, A. Tchekhovskoy$^2$, A. Ingram$^3$, M. van der Klis$^1$\\
  $^{1}$Anton Pannekoek Institute for Astronomy, University of Amsterdam, Science Park 904, 1098 XH Amsterdam, The Netherlands\\
  $^{2}$Center for Interdisciplinary Exploration \& Research in Astrophysics (CIERA),
  Physics \& Astronomy, Northwestern University, Evanston, IL 60208, USA\\
  $^{3}$ Department of Physics, Astrophysics, University of Oxford, Denys Wilkinson Building, Keble Road, Oxford, OX1 3RH, UK
}

\begin{document}

\date{Accepted. Received; in original form}
\pagerange{\pageref{firstpage}--\pageref{lastpage}} \pubyear{2015}

\maketitle

\label{firstpage}

\begin{abstract}
  Prevalent around luminous accreting black holes, thin discs are challenging to resolve in numerical simulations.
  When the disc and black hole angular momentum vectors are
  misaligned, the challenge becomes extreme, requiring adaptive meshes to
  follow the disc proper as it moves through the computational grid.
  With our new high-performance general relativistic
  magnetohydrodynamic (GRMHD) code H-AMR we have simulated the
  thinnest accretion disc to date, of aspect ratio
  $H/R\approx0.03\approx1.7^\circ$ , around a rapidly spinning ($a=0.9375$) black hole, using a
  cooling function. Initially tilted at
  $10^\circ$, the disc warps inside $\sim 5r_g$ into alignment with
  the black hole, where $r_g$ is the gravitational radius. This is the
  first demonstration of Bardeen-Petterson alignment in MHD with
  viscosity self-consistently generated by magnetized turbulence. The
  disc develops a low-density high-viscosity ($\alpha_{\rm eff}\sim1.0$)
  magnetic-pressure--dominated inner region at $r\lesssim25r_g$ that
  rapidly empties itself into the black hole.
  This inner region may in reality, due to thermal decoupling of ions and electrons, evaporate into a radiatively inefficient accretion flow if, as we propose, the cooling time exceeds the accretion time set by the order unity effective viscosity.
  We furthermore find the unexpected result that even our very thin disc can
  sustain large-scale vertical magnetic flux on the black hole, which
  launches powerful relativistic jets that carry $20{-}50\%$ of the
  accretion power along the angular momentum vector of the outer
  tilted disc, providing a potential explanation for the origin of
  jets in radio-loud quasars. 

\end{abstract}


\section{Introduction}
\label{sec:introduction}
Black holes (BHs) in X-Ray binaries (XRB) and possibly active galactic nuclei (AGN) cycle during their lifetimes through different accretion states
characterized by the total luminosity and spectral hardness. It is
widely believed that the accretion luminosity expressed as a fraction of the
Eddington limit $L_{\rm Edd}$ is an important factor in determining the BH's
accretion state \citep{Esin1997,Remillard2006,McClintock2006}.
At low luminosities ($L\lesssim0.01 L_{\rm Edd}$), in the low-hard
state, a promising model is the \textit{advection dominated
  accretion flow} (ADAF, \citealt{Ichimaru1977,Narayan1994,Narayan1995A,Narayan1995B}). In an ADAF, the disc surface density is so low that the
plasma can decouple into a two-temperature electron-ion plasma
\citep{Shapiro1976}. Since the ions are unable to cool on the accretion time, most
of the dissipated energy is advected
inwards in the disc or expelled in outflows leading to a low radiative efficiency. It is also known through general
relativistic magnetohydrodynamic (GRMHD) simulations that these thick
accretion discs can sustain and advect inwards large scale poloidal
magnetic flux (e.g. \citealt{Villiers2003, Mckinney2006,Beckwith2008, mckinney2009,Tchekhovskoy2011,Tchekhovskoy2012,Mckinney2012}) which launches powerful jets
when it reaches the central BH \citep{bz77}. The large scale height
makes ADAFs numerically easy to study since they do not require high
resolutions and have short viscous times.
They have been studied extensively in GRMHD and are relatively well
understood.

However, the observed emission from X-ray binaries in the
high-soft state, and from high-Eddington fraction AGN, is
incompatible with the ADAF solution. Their thermal emission spectrum
requires a geometrically thin, optically thick accretion disc
\citep{ss73,Novikov1973}. Though X-ray binaries and AGN only spend a rather short
amount of time in the thin disc regime, most of the BH growth
and feedback may still occur there since the accretion rate is several
orders of magnitude higher. Indeed, a typical bright quasar (or XRB) radiates at $\sim10\%$ of the Eddington rate,
$\dot M_{\rm Edd} = 10L_{\rm Edd}/c^2$, and its disc thickness is
extremely small,
$H/R\simeq 0.01\times (\dot M/0.1\dot M_{\rm Edd})^{0.9}$ 
\citep[see
e.g.\ Fig.~3 in][]{2015MNRAS.453..157P}. Thus, it is crucial to study these systems
to understand the growth and feedback of supermassive BHs.

Modelling of relativistic iron line profiles suggests that most bright
local AGN harbour rapidly spinning supermassive BHs, with spin parameter
$a>0.8$ \citep{Reynolds2014}. Because the infalling material is
unaware of the orientation of the BH spin, its angular
momentum vector is expected to be misaligned with respect to the black
hole spin vector, resulting in a tilted accretion disc. In fact, there exist several such candidates for both X-ray binaries
\citep{Hjelming1995,Greene2001,Maccarone2002} and AGN
\citep{Caproni2006,Caproni2007}.  Analytic theory predicts that the
inner parts of a thin disc would align with the BH midplane
due to the Bardeen and Petterson effect (hereafter BP,
\citealt{bp75,Papaloizou1983,Kumar1985,Pringle1992,Ogilvie1999}) out to a
Bardeen-Petterson radius, $r_{\rm BP}$. The alignment can have
profound consequences for the growth and feedback of supermassive BHs
\citep{Rees1978, Scheuer1996, Natarajan1998} since a larger $r_{\rm BP}$ implies a
larger alignment torque on the BH, forcing it into alignment
with the outer disc on very short timescales. Aligned accretion subsequently leads to rapid BH spinup by the disc, even if most of the infalling material is initially misaligned (and even counter-aligned in some cases, see \citealt{King2005}). 

The BP effect is caused by the interplay between general relativistic
Lense-Thirring precession \citep{LT1918} and viscosity. BP alignment has been
observed in pioneering smoothed particle hydrodynamics (SPH)
simulations \citep{Nelson2000,Lodato2007,Lodato2010}. However, because these
simulations are non-relativistic and hydrodynamic, they are unable to include the full effects of GR and treat the anisotropy of magnetized turbulence accurately. In fact, GRMHD is a powerful way to
model the non-linear nature of the magnetorotational instability
(MRI) driven turbulence \citep{Balbus1991} responsible for the viscosity
whilst including the full effects of GR. However, GRMHD work at a
moderate thickness of $H/R=0.08$ did not
find Bardeen-Petterson alignment \citep{Texeira2014,Zhuravlev2014}. Moreover, there
is growing evidence that the interaction between the disc, magnetized
corona and the jets should be taken into account. Jets
can torque the inner accretion disc into alignment \citep{Mckinney2013} before they align with the outer accretion flow \citep{Liska2018A}. The corona, which we define as the hot bloated magnetic pressure supported flow surrounding the thin disc, on the other
hand, is not expected to align since thick flows cannot exhibit Bardeen-Petterson
alignment \citep{Ivanov1997, Papaloizou1983}. Thus, full GRMHD simulations, which describe the entire
\emph{thin}-disc--corona--jet system, are uniquely positioned to address the more than $40$
year old fundamental problem whether BP alignment occurs in the inner parts of thin tilted discs.

Indications are that jets in systems that contain a thin disc are rare:
only $10$\% of quasars, or luminous AGN, are observed to produce
relativistic jets and the associated radio emission (e.g
\citealt{Sikora2007}), while there are no convincing observations in soft state
X-ray binaries (though see \citealt{Rushton2011}). It is crucial to
understand what factors are responsible for the formation and
destruction of jets since they can be the dominant feedback mode in
galaxy clusters (see e.g. \citealt{fabian2012}). Early theoretical work suggested that thin accretion discs are not expected to have jets since the
poloidal magnetic flux may diffuse out before it can advect inwards
\citep{Lubow1994}. However, the non-uniform vertical structure of accretion discs and their
turbulence may aid in the inward advection of poloidal magnetic flux \citep{Rothstein2008,Giulet2012,Giulet2013}.

The small vertical extent of thin discs makes them very difficult to
study numerically, with the computational cost scaling as $(H/R)^{-5}$
per accretion time. Because of this high cost, numerical work studying the
physics of such discs has been mostly limited to shearing box
simulations and semi-analytical studies.  In fact, there are no 3D
GRMHD simulations available for thin discs of aspect ratios
$H/R<0.05$, and the thinnest discs so far simulated in 3D GRMHD ($0.05<H/R<0.1$) were aligned, which enabled the vertical wavelength of the MRI to be resolved by a grid focused on the equatorial plane, leading to cells
compressed in the $\theta$-direction and elongated in the $r$- and
$\phi$-directions (e.g.
\citealt{Shaffee2008, Noble2009,Noble2010, Penna2010,Texeira2014,Avara2016,Texeira2017}). 
Studying tilted discs,
whose orbital motion does not conform to the main directions of the
grid, is more difficult than aligned ones because one can assume
neither axisymmetry nor use elongated cells to speed up the
simulations.

In this work we present the thinnest global GRMHD accretion disc
simulations to date and study Bardeen-Petterson alignment and
jet launching. %
We describe our setup in Sec.~\ref{sec:numerical-models},
present the results and discussion in Secs.~\ref{sec:results} and
\ref{sec:Discussion}, and conclude in Sec.~\ref{sec:conclusions}.

\begin{figure*}
\centering
\includegraphics[width=7.0in,trim=0cm 0cm 0cm 0cm]{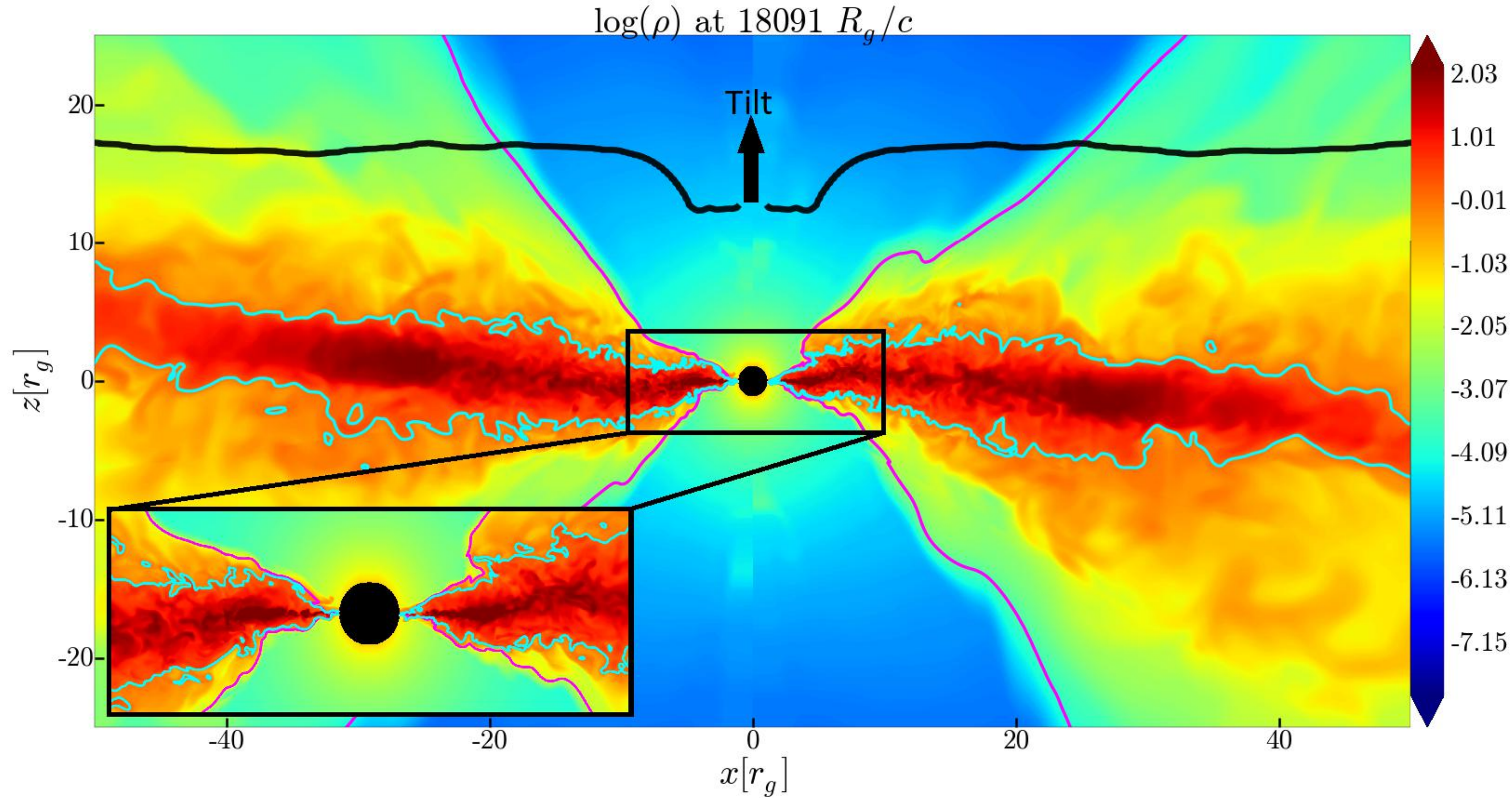}
\caption{Vertical slice at $t = 18,\!091 r_g/c$ through density of our tilted disc with an
  aspect ratio $H/R=0.03$, the thinnest to date in a GRMHD simulation,
  shows that the inner parts of the disc ($r\lesssim 5r_g$) align with
  the rapidly spinning BH's equator (horizontal in the figure; see the
  inset for a zoom-in). The warp is accompanied by a $\sim 180^\circ$
  sweep in the precession angle (see Fig.~\ref{fig:tiltvsr}b),
  creating a `negative' tilt around $r\sim 8 r_g$. Thick black curve shows that the disc tilt
  $\mathcal T$ smoothly decreases toward small radii before flattening at
  $0^\circ$ for $r\lesssim 5r_g$ (see also Fig.~\ref{fig:tiltvsr}a). This is the first
  demonstration of \citet{bp75} alignment in a GRMHD simulation, i.e.,
  in GR and in the presence of large- and small-scale turbulent
  magnetic stresses.
  Magenta lines show the corona-jet boundary, $p_{\rm b}=1.5\rho c^2$,
  and cyan lines the disc-corona boundary, $\rho = 1.2$.}
\label{fig:bp}
\end{figure*}

\section{Numerical models}
\label{sec:numerical-models}
We use for this work our state-of-the-art GPU-accelerated GRMHD code
H-AMR \citep{Liska2018A}, which finds its heritage in the HARM2D code
\citep{Gammie2003,Noble2006} but has been significantly %
expanded %
to take advantage of vectorization and SIMD instructions, %
achieving $1.2\times
10^6$ zone-cycles per second on a single Intel Skylake 3.3GHz CPU
core, and include advanced features discussed below. We have developed
a CUDA version of the code that reaches $10^8$ zone-cycles/s on an NVIDIA Tesla
V100 GPU. It also features a
staggered grid for constrained transport of magnetic fields
\citep{Gardiner2005}, uses an Harten-Lax-van Leer (HLL) Riemann solver
\citep{Harten1983} and advanced features such as adaptive mesh
refinement (AMR) and local adaptive time-stepping (LAT) that bring
down the cost of the simulation described in this work by $2$ extra
orders of magnitude (in comparison to using a uniform grid and a
global, fixed timestep).

Our thin disc model considers a spinning BH of spin parameter
$a=0.9375$. We insert an initial Fishbone \& Moncrief torus
\citep{Fishbone1976}, which is maintained in hydrostatic equilibrium
by the BH's vertical component of gravity counterbalancing the
pressure forces in the disc. This torus has an inner radius
$r_{\rm in}=12.5r_g$, where $r_g=GM_{\rm BH}/c^2$ is the gravitational radius, with the pressure maximum at $r_{\rm
  max}=25r_g$. We use an ideal gas law equation of state,
$p_{\rm g}=(\Gamma-1)u_{\rm g}$, with $u_{\rm g}$ the gas internal
energy density, $p_g$ the gas pressure
and $\Gamma=5/3$ the adiabatic index. We insert a poloidal magnetic
field in the torus described by a covariant vector potential
$A_{\phi}=(\rho-0.05)^{2}r^{3}$, with $\rho$ the gas density of the
torus. The magnetic field is subsequently normalized by setting
$\Bar{\beta}={\max p_{\rm g}}/{\max p_{\rm b}}=30$, where $p_{\rm b}$ is
the magnetic pressure and both maxima are taken over the torus separately. The disc is tilted
by $\mathcal T_0 = 10^\circ$ with respect to the BH equator (see
\citealt{Liska2018A} for details). Since the code is scale-free, we set the \textit{initial} gas density maximum to $\rho_{\rm max}=1$.
To maintain the desired value of disc thickness, $(H/R)_{\rm target}=0.03$, we cool
the disc towards its target temperature on the Keplerian timescale using a prescribed source term
(\citealt{Noble2009}). We disable this cooling
function when $p_b/(\rho c^2)\gtrsim5$ at $r\gtrsim 10 r_g$ to avoid cooling the jets.

In this work we for the first time use the full AMR capability of H-AMR, in order to ensure sufficient
resolution within the thin disc. For the refinement criterion we use a density cutoff equal to $\sim 4$ percent of the maximum disc density. The lower refinement levels are set such that jumps in spatial resolution are limited to a factor $2$. To avoid noise from sporadic refinement and derefinement, we
only derefine at a two times lower density than set as the refinement criterion. Typically, we use $3$ levels of AMR and attain a speedup by a
factor 32--60 in comparison to an equivalent uniform grid. 
By evolving lower AMR levels or parts of the grid further from the black hole, which have larger cell sizes, at a larger timestep, LAT gives an additional speed-up of
factor $\sim 5$ while reducing inversion errors in relativistic regions (Chatterjee et al 2018, in prep).
 The effective
resolution of $2880\times864\times1200$ cells in \hbox{$r$-,} \hbox{$\theta$-,} and
$\phi$-directions, respectively, resolves the target disc
thickness, $(H/R)_{\rm target}= 0.03$, by approximately $8$ cells in all 3 dimensions, and the base grid of
$720\times216\times300$ cells guarantees that the jets and corona are
also sufficiently resolved. As we will see in Sec.~\ref{sec:radialstructure}, the fastest growing
MRI wavelength is resolved by $\gtrsim10$ cells over most of the disc. 

Operating in Kerr-Schild spherical polar coordinates, we place the inner $r$-boundary just inside the event horizon and the
outer $r$-boundary at $10^5r_g$, so that the flow is unaffected by the
boundaries. We use transmissive polar boundary conditions
in the $\theta$-direction \citep{Liska2018A}, and periodic
boundary conditions in the $\phi$-direction.

In BH powered jets, the gas either drains off the field lines into the
BH or gets flung away along the field lines into the jet. This leads
to a runaway drop in density around the jet's stagnation surface, at
which the outflow velocity vanishes. To avoid the development of vacuum regions
and the breakdown of ideal MHD, we replenish the density in the
regions where the density drops too low. For this, we follow the approach of
\citet{Ressler2017} and approximate physical processes that mass-load
relativistic jets at their base by applying a density floor of
$\rho_{\rm floor}c^2 = p_{\rm b}/10$ throughout the jet. This adds a small
amount of density on the field lines and does not noticeably affect
the energetics of the jets.

\section{Results}
\label{sec:results}
We start our analysis at $t=10^4r_g/c$
when the disc has cooled to its equilibrium thickness, reaching a highly turbulent
state, as seen in 
Figure~\ref{fig:bp}. We measure throughout
this work vector quantities for the disc, corona and jet in tilted
spherical polar coordinates $r,\tilde{\theta},\tilde{\phi}$ that are aligned with the
disc's angular momentum at each radius. We define the jet-corona
boundary at $p_{\rm b}=1.5\rho c^2$ and the corona-disc boundary at
$\rho=1.2$; this is approximately $10^{-2}$ times the maximum
density in the disc after the initial cooling completes.

\subsection{Bardeen-Petterson Alignment}
\label{sec:bp}
Figure~\ref{fig:bp} shows that the inner parts of the accretion disc ($r\simless5 r_g$)
align with the BH equator. This is the first demonstration of BP alignment in GRMHD. This alignment happens relatively soon, already $\sim(500-1000)r_g/c$ after accretion starts, and persists throughout the simulation. The outer parts remain
tilted, and the disc develops a smooth warp in between. 

We can see this
more quantitatively by introducing the tilt, $\mathcal{T}$, and
precession, $\mathcal{P}$, angles. We define them as polar and azimuthal angles between
the disc's angular momentum and the BH spin vector
(e.g. \citealt{fragile2005,Fragile2007}). While this
definition works well for the disc and the corona, it produces large
fluctuations when applied to jets since they can undergo strong kinks and
pinches, which violently change the jet angular momentum direction (by
more than $10^\circ$) on very short time and length scales. For this
reason, we adopt a more robust way of calculating the tilt and
precession angles for the jets, as described in Appendix~B of
\citet{Liska2018A}. Namely, instead of considering the angular
momentum vector orientation, we calculate the jet tilt and
precession angles at each radius from the jet's average position, ($X,Y,Z$), weighted by the
magnetic pressure $p_b$:
\begin{equation}
\epsilon=p_{\rm b}\times \Heavi\left({p_{\rm b}}-1.5 {\rho c^2}\right),
\label{eqn:epsilon}
\end{equation}
where $\Heavi$ is the Heaviside step function that zeroes out the weight
outside of the magnetized jet body.

\begin{figure}
\centering
\includegraphics[width=\linewidth,trim=0cm 0cm 0cm 0cm,clip]{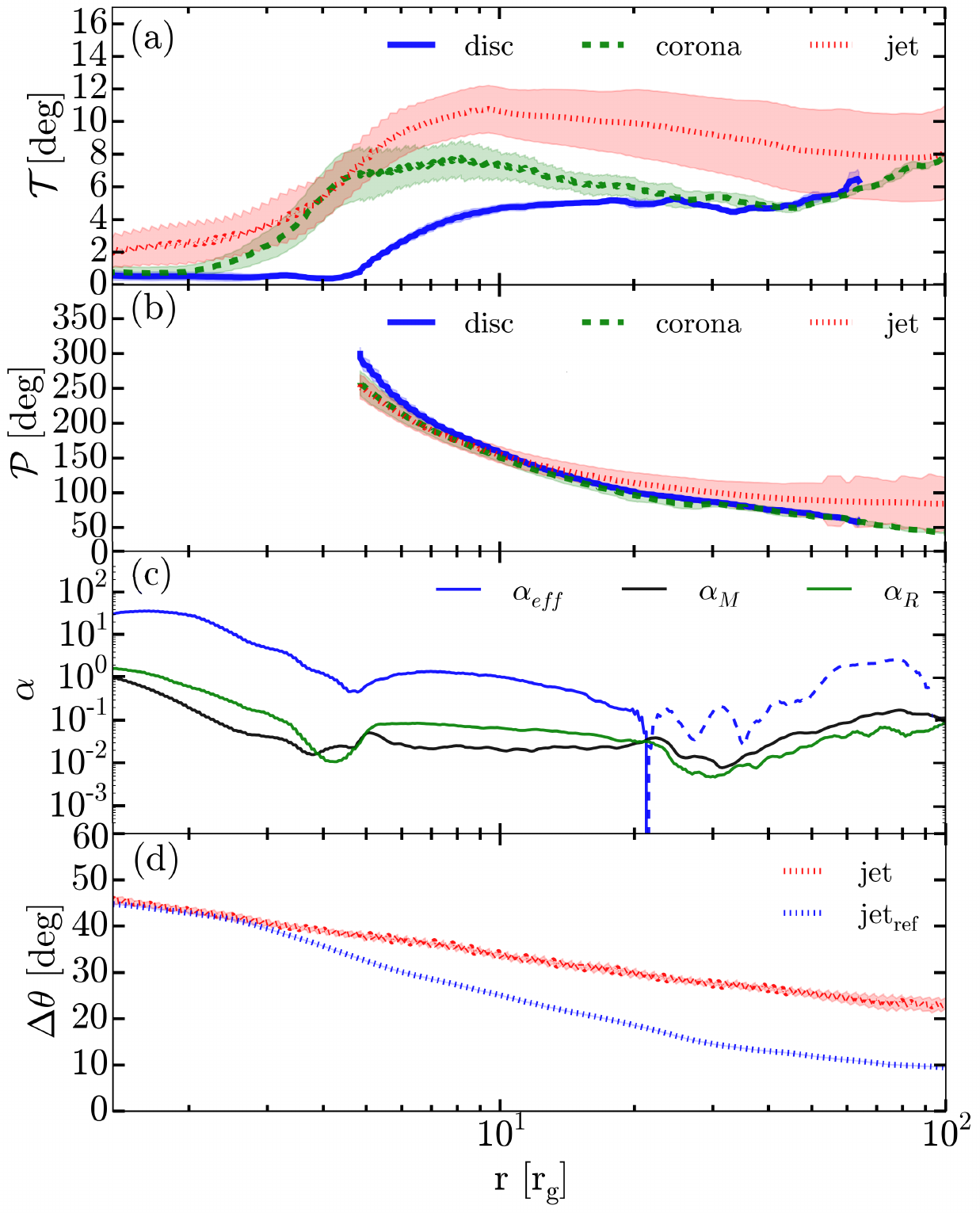}
\vspace{-10pt}
\caption{ Radial profiles of quantities in our thin disc system, averaged
  over the time interval, $2 \times 10^4<t<2.25 \times 10^4r_g/c$. The shaded regions represent the time variability over this interval.  (a)
  Tilt angles of the disc, corona and jet. Whereas the outer disc is tilted by
  $\mathcal T\approx 5^\circ$, the inner disc at $r\lesssim 5r_g$ is
  aligned with the BH equator, indicating the presence of \citet{bp75}
  alignment. While the disc aligns rapidly near the BH, the corona and
  jet show a single radial tilt oscillation peaking at $r\sim10 r_g$,
  which is characteristic behaviour of thick accretion discs that do
  not show Bardeen-Petterson alignment
  (e.g., \citealt{Fragile2007, Liska2018A}). That the corona and the jets
  at smaller radii
  align less readily with the BH than the disc, suggests that the alignment is driven
  by the disc dynamics rather than that of the corona or jet. (b) The precession angle for the disc, corona, and
  jet are roughly consistent with each other at large radii,
  suggesting they become co-aligned. As the disc
  aligns at small radii, $r\lesssim 5r_g$, the precession angles become
  ill-defined, and we do not show them there. (c) The effective viscosity and the sum of the Maxwell and Reynolds stresses exceed the disc's scale height ($h/r\sim0.03$), suggesting our disc is in the viscosity dominated warp propagation regime (see Secs.~\ref{sec:radialstructure} and \ref{sec:disc-bp}) (d) The half-opening angle of our jets (red curve) exceeds
  the opening angle of the jets in a thick $H/R \sim 0.3$ disc model \citep[blue dotted
  curve,][]{Liska2018A}, suggesting that a smaller disc thickness in
  this work provides less pressure support for the jet and causes
  its opening angle to widen. }
\label{fig:tiltvsr}
\end{figure}

Figure~\ref{fig:tiltvsr}(a) shows that the inner parts of the
accretion disc, at $r\lesssim 5r_g$, are well-aligned with the BH
equator: 
The tilt angle $\mathcal T^{\rm inner}_{\rm disc}\ll 1^\circ$ of the
inner disc ($r\lesssim 5 r_g$) is much smaller than the tilt angle
$\mathcal T^{\rm outer}_{\rm disc}\gtrsim 5^\circ$ of the outer disc
($r\gtrsim10r_g$).  The alignment radius remains steady at $r_{\rm BP}
\approx 5 r_g$ during the course of our
simulation. Note that $\mathcal
T^{\rm outer}_{\rm disc}\simeq 5^\circ$ is smaller than the initial
tilt of the disc,
$\mathcal T_0 = 10^\circ$. This is because the disc as a whole undergoes global
alignment, which is distinctly different than the BP effect (Sec.~\ref{sec:global-evolution}).

Since the corona is relatively thick, spanning an opening angle of
around $30-70^\circ$ (e.g. Figure~\ref{fig:bp}), it is not
expected to show BP alignment (Sec.~\ref{sec:introduction}).
For such thick structures as the corona, the analytic theory predicts that BP
alignment is suppressed and accretion instead occurs in a misaligned fashion
through radial tilt oscillations
(e.g. \citealt{Ivanov1997,Lubow2002}). Indeed,
Fig.~\ref{fig:tiltvsr}(a) shows a single radial
tilt oscillation in the corona and jet peaking around $r\sim10r_g$. However, the corona still manages to align within $r\lesssim 2.5r_g$, possibly due to torque from the disc. Since the angular momentum of the corona is negligible, the disc can easily affects its alignment.

While the inner jets are relatively closely aligned with the BH, the outer
jets are torqued into misalignment by the corona. This is consistent
with the previous work \citep{Liska2018A}, which in the context of
thick $H/R\sim0.3$ discs found that the outer disc-corona system is responsible for
reorienting and collimating the jets. Similarly the precession angle
of the disc, corona and jet are closely related at large radii, as
seen in Fig.~\ref{fig:tiltvsr}(b). At small radii, as the system
becomes aligned, the precession angle becomes ill-defined, and we do
not show it.

Since the disc shows much ($\sim2\times$) better alignment with the
BH than the jet and corona, it is unlikely that the jet can be
responsible for torquing the inner disc into (partial) alignment as has been demonstrated for
thicker discs (e.g., \citealt{Mckinney2013}). If this were the case, one would expect the jet tilt to be smaller than the disc tilt and, since the jet actually has to transmit its torque through the corona to the disc, one would expect better alignment for the corona as well.

\subsection{Global precession and alignment}
\label{sec:glob-prec-alignm}

\begin{figure}
\centering
\includegraphics[width=\linewidth,trim=0cm 0cm 0cm 0cm,clip]{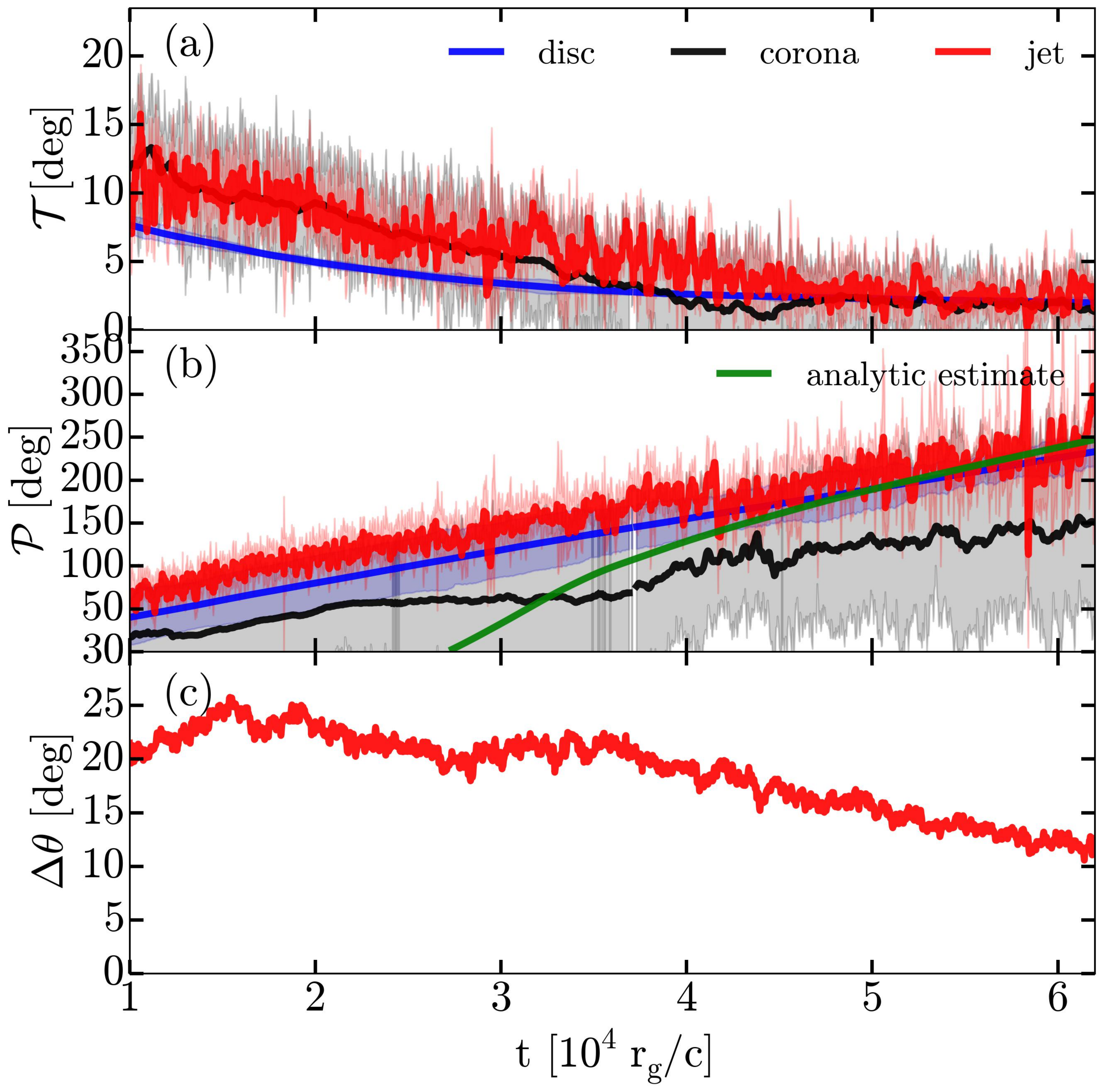}
\vspace{-10pt}
\caption{The
  tilt angle (a) and
  precession angle (b), as
  functions of time for the entire disc (blue), coronal wind within
  $r<250r_g$ (black) and jet (between $10r_g<r<100r_g$, red). The shaded regions show the standard
  deviations within $10r_g<r<40r_g$. The whole system precesses and over time aligns with the
  BH. The analytic estimate for the precession rate (green) is only valid at later times. (c) The jet
  half-opening angle at $r=200r_g$ decreases as function of time, possibly due to a fall in jet power
  (Figure~\ref{fig:timeplot}a,b).}
\label{fig:tiltvst}
\end{figure}

In addition to exhibiting the BP effect, which depends on space and is independent of time
and in which the inner part of the disc aligns with the BH, the entire
accretion system additionally undergoes global alignment that depends
on time and is independent of radius. Figure \ref{fig:tiltvst}(a),(b)
shows the time evolution of the average tilt, $\mathcal{T}$, and
precession, $\mathcal{P}$, angles for the disc, jet and
corona. The disc and corona are angular momentum averaged, the jet is magnetic pressure averaged (Sec. \ref{sec:bp}). Similar to thick $H/R \sim 0.3$ discs (\citealt{Liska2018A, Liska2019A}), our thin disc aligns \emph{as a whole} with the BH spin. \citet{Sorathia2013} proposed that alignment (BP alignment or global alignment) may be caused
by the turbulent mixing between disc annuli with different precession angles (Fig.~\ref{fig:tiltvsr}b shows that the precession angle decreases as function of radius), which leads to cancellation of misaligned angular momentum and thus produces net alignment. %

As we discussed above, this is \emph{not}
BP alignment for two reasons. First, BP
alignment is characterized by a steady state solution that is
established on a timescale shorter than an accretion time of the disc
(our disc lost $\sim 20 \%$ of its initial mass by the end of this
simulation): otherwise, most of the disc's mass would accrete
misaligned. %
Second, if the disc is fed externally, such as by a
larger thin disc or by fallback material in a tidal disruption event,
such global alignment may disappear as the inner disc is replenished
by the gas carrying the misaligned angular momentum on the accretion
time of the inner disc. Indeed, \citet{Liska2018A} showed that
global alignment becomes slower as the disc size becomes larger due to viscous spreading. 

The precession period of around $10^5r_g/c$ is consistent with a
Type-C quasi-periodic oscillation (QPO) frequency of $0.2\rm{Hz}$ for
a $10M_\odot$ BH (e.g. \citealt{2016MNRAS.461.1967I}). %
Although Type-C QPOs are not observed for the clean thin discs thought to be present in the soft state, we note that the precessing corona in our simulation could give rise to the QPOs typically observed in the Comptonised radiation from X-ray binaries during the transition from hard to soft state. Note that the
precession angle of the corona is smaller than that of the disc, as
seen in Fig.~\ref{fig:tiltvst}(b). Therefore, the precession of the
corona lags that of the disc. Thus, the hard variations might lag the
soft variations (see \citealt{Liska2019A}).
Note that, due to global alignment, which
causes the outer disc to align with the black hole to within $2^\circ$
by $6\times10^4 r_g/c$ (Fig.~\ref{fig:tiltvst}a), the precession
cannot be sustained for more than a single period. 
\begin{figure}
\centering
\includegraphics[width=\linewidth,trim=0cm 0cm 0cm 0cm,clip]{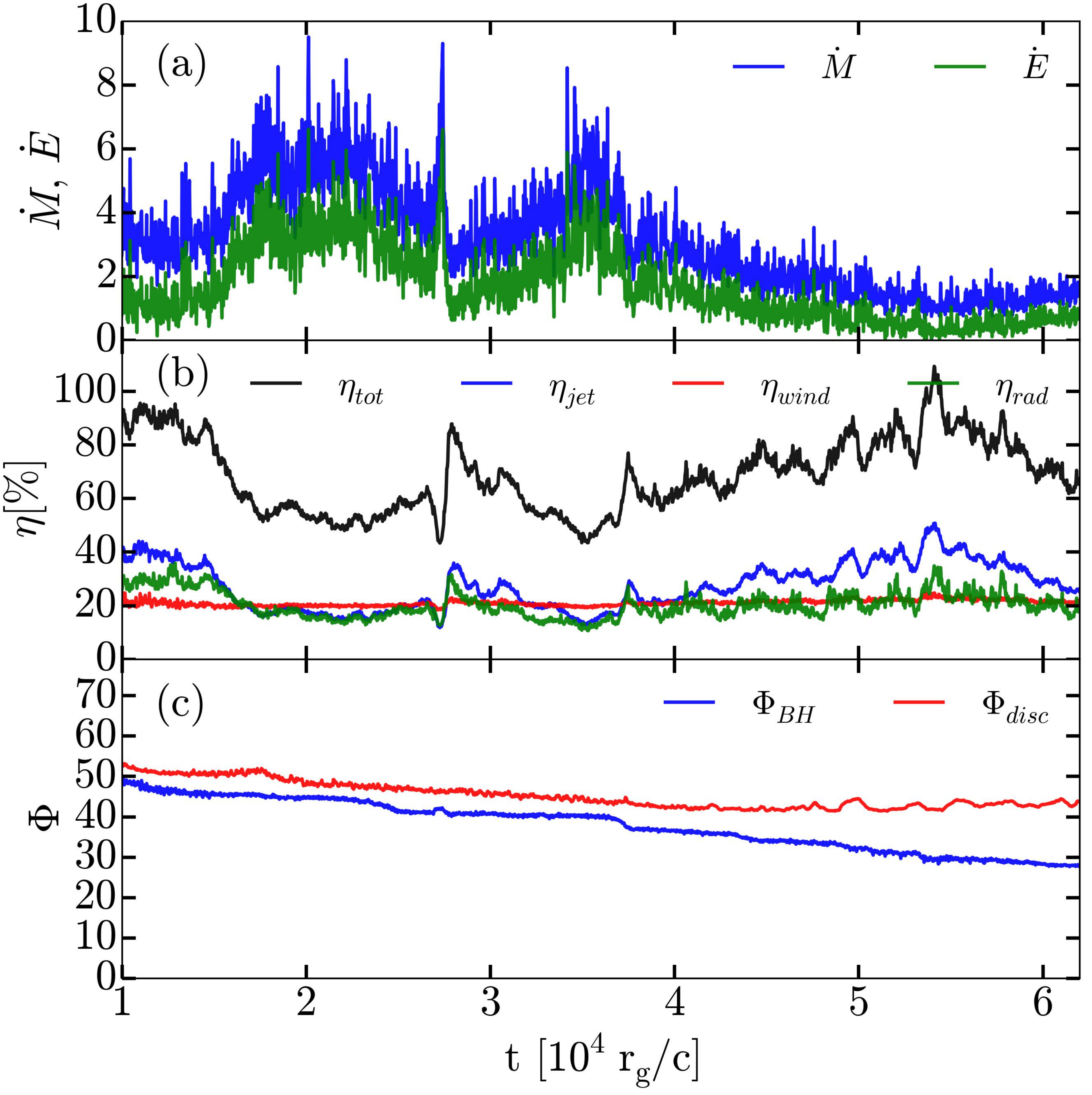}
\caption{Time-dependence of various quantities in our
  simulation. (a) Mass $\dot M$ and energy $\dot E$ accretion rates, which
  initially oscillate but stabilize later on. (b) The total
  outflow efficiency (black) and its constituents: jets (blue), coronal wind
  (red), and radiative (green) contributions. The total efficiency of
  $60{-}80\%$ is unprecedentedly high for such a thin disc as ours and
  exceeds the standard thin disc efficiency,
  $17.9\%$ \citep{Novikov1973}, by a factor of $3$. Particularly surprising is the high
  jet efficiency, $\eta_{\rm jet}\sim 20\%$, indicating that thin
  discs are capable of producing powerful relativistic jets. (c) The magnetic flux on the BH
  (blue) and in the disc (blue). While the disc magnetic flux,
  $\Phi_{\rm disc}$ remains roughly constant at late times,
  $t\gtrsim4\times10^4r_g/c$, the flux on the BH decreases, suggesting
  that the BH might be leaking its flux into the disc and that the jet
  formation by our thin disc might be a transient phenomenon
  reminiscent of transient jets in XRBs (see
  Sec.~\ref{sec:Discussion}).}
\label{fig:timeplot}
\end{figure}

We can analytically estimate the precession period
$P_{\rm LT}$ by calculating the total perpendicular angular momentum
$L_{\bot}$ and Lense-Thirring torque $\tau_{\bot}$ directly from the
stress energy tensor $T^{\mu}_{\nu}$ and test-particle LT precession
rate $\Omega_{\rm LT}$:
\begin{align}
L_{\bot}&=\iiint_{r_{\rm in}}^{r_{\rm max}}T^{t}_{\tilde{\phi}}\sin\mathcal{T}\, dV,\\
\Omega_{\rm LT}&=\frac{1}{r^{3/2}+|a|}\left(1-\sqrt{1-4\frac{|a|}{r^{3/2}}+3\frac{a^2}{r^2}}\right),\\
\tau_{\bot}&=\iiint_{r_{\rm in}}^{r_{\rm max}}\Omega_{LT} \times T^{t}_{\tilde{\phi}}\sin\mathcal{T}\,dV,\\
  P_{\rm LT}&=2 \pi \times \frac{L_{\bot}}{\tau_{\bot}}.
              \label{eq:prec}
\end{align}
Figure~\ref{fig:tiltvst}(b) shows that the above analytic expression
for the precession period approximately agrees with the simulation at
later times if we assume $r_{\rm in}=1.5\times r_{\rm isco}$. However,
at earlier times this overestimates the precession rate. To remedy
this, $r_{\rm in}$ would need to increase in time, however it is
unclear why this would happen.

\begin{figure*}
\centering
\includegraphics[width=6.5in,trim=0cm 0cm 0cm 0cm,clip]{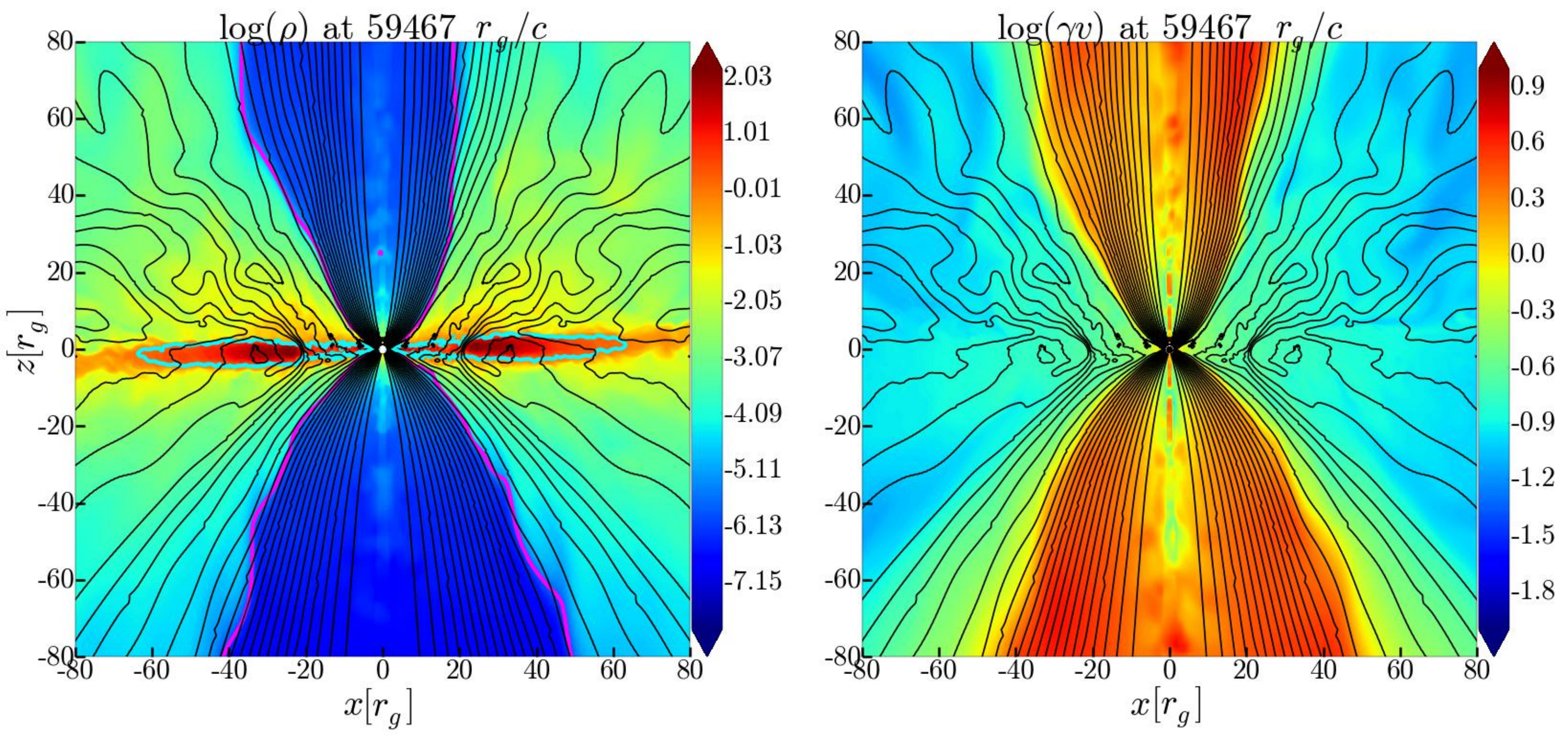}
\vspace{0pt}
\caption{Transverse slices of logarithm of density $\rho$ and the logarithm of the magnitude of the
  proper velocity $\gamma v$, at $t\approx6\times10^4r_g/c$. The field lines (black) that connect to the BH launch 
  relativistic jets (demarcated by magenta lines at $p_{\rm b}=1.5\rho c^2$), while field lines threading the disc 
  give rise to a non-relativistic coronal wind (disc-corona boundary is shown with the cyan line at $\rho=1.2$, see Sec.~\ref{sec:results}). The twin low-density polar jets accelerate
  to $\gamma v\sim 5c$, whereas the winds are limited to at most $\gamma
  v\lesssim c$, or $v\lesssim0.7c$, which occurs near the edges of the jets.
}
\label{fig:contourplot}
\end{figure*}

\subsection{Outflow and radiative efficiency}
\label{sec:global-evolution}
Figure~\ref{fig:timeplot}(a) shows that even after the disc has
approached the equilibrium thickness at $t\sim10^4r_g/c$, both the
mass $\dot{M}$ and energy $\dot{E}$ accretion rates at the event
horizon, $r_{\rm H}=r_g[1+(1-a^2)^{1/2}] = 1.35r_g$, show several
peaks. They stabilise to within a factor of a few at
$t \gtrsim 4 \times 10^4r_g/c$, suggesting that by then the disc has reached a
quasi-steady state. 
That $\dot M c^2> \dot E$ shows that the BH
accretes more energy than it ejects out.
To quantify this, we introduce dimensionless energy outflow efficiencies
for the relativistic jets,
\begin{equation}
\eta_{\rm jet}=-\frac{\dot{E}_{jet}}{\langle \dot{M}c^2 \rangle_{t}},
\end{equation}
where $\dot E_{jet}$ is the energy accretion rate in the jets at
$r=20r_g$ (chosen to be larger than $r_{\rm H}$ for robustness of
identification of the jets), and the disc winds,
\begin{equation}
\eta_{\rm wind}=\frac{\langle \dot{M}c^2 \rangle_{t}-\dot{\langle E \rangle_{t}}+\dot E_{jet}}{\langle \dot{M}c^2\rangle_{t}}.
\end{equation}
In the above, e.g. $\langle \dot{M} \rangle_{t}$ is
the running time-average of the mass accretion rate over an interval
of $\pm500r_g/c$. Note that, from these equations, any of the rest mass energy of disc material not accreted onto the BH horizon, and not ejected in the jets, is by definition lost in a wind. Figure~\ref{fig:timeplot}(b) shows that the jet
efficiency resides in the range $\eta_{\rm jet} = 20{-}50\%$, implying
that as much as $50\%$ of the mass-energy accreted by the BH can be
extracted by large-scale magnetic flux from the black hole spin energy
and carried out in the form of
Poynting-flux dominated jets. Figure~\ref{fig:contourplot} shows that
the field lines responsible for the jet launching are anchored in
the BH: they extract the energy via the \citet{bz77} effect.
That the jets are so efficient is a surprising result: the standard
expectation is that thin discs are incapable of holding on to
large-scale poloidal magnetic flux and hence are not expected to have
powerful jets \citep{Lubow1994}.

In addition to the jets, the disc also launches a sub-relativistic
disc wind, which is magnetic pressure dominated and thus similar to a corona, with energy efficiency $\eta_{wind}\sim
20\%$. The field lines threading this outflow are anchored in the
disc, suggesting it may be magneto-centrifugally driven by the
\citet{bp82} mechanism. However, the precise nature of the launching
mechanism is not entirely clear, since time-dependence, (magnetic)
pressure gradient forces, and buoyancy forces, as well as energy
extraction from the BH, may also play significant roles in launching
this outflow. The outflow has a radial velocity of around $0.02-0.2c$
between $10r_g<r<500r_g$. This compares favorably to AGN ultra-fast
outflows, whose velocities lie in the range of $0.1{-}0.4c$
\citep[e.g.,][]{2010A&A...521A..57T,2011ApJ...742...44T}, but is on the (very) high
end of winds detected in the soft state of XRBs \citep[e.g.,][]{Ponti2012,2016AN....337..512P,2016ApJ...822L..18M}. 

To characterize the radial distribution of the magnetic flux, we
define a 1D
poloidal magnetic flux function, $\Phi(B,r)$, as the
maximum of the poloidal magnetic flux at each radius of magnetic field~$B$, 
\begin{equation}
\Phi(B, r)=\max_{\tilde\theta_{\rm m}}\iint_{0}^{\tilde\theta_{\rm m}} B(r,\tilde\theta,\tilde\phi)dA_{\tilde\theta\tilde\phi},
\label{eq:PhiBr}
\end{equation}
where $dA_{\tilde\theta\tilde\phi} = \sqrt{-g}d\tilde\theta d\tilde\phi$ is the area element 
and $g$ is the metric determinant. This gives us the
magnetic flux on the BH, $\Phi_{\rm BH} = 0.5\Phi(|B_r|,r_{\rm H})$, and the
the magnetic flux content of the disc, $\Phi_{\rm
  disc}=\max_r\Phi(B_r,r)$. Figure~\ref{fig:timeplot}(c) shows that at
early times $\Phi_{\rm BH} \simeq \Phi_{\rm disc}$, implying that most
of the positively-oriented magnetic flux resides on the BH and very
little in the disc. Over time, both initially decline roughly in the same
proportion, but at $t \gtrsim 4 \times 10^4r_g/c$ the value of
$\Phi_{\rm disc}$ flattens out whereas $\Phi_{\rm BH}$ continues to
decline. This suggests that the magnetic flux diffuses out of the BH
and into the disc. The stability of $\Phi_{\rm BH}$ on short
timescales suggests that the disc is not in the magnetically arrested
disc regime \citep[MAD,][]{Narayan2003,Igumenschev2003}. However, it
is not yet known whether the magnetic flux expulsions and associated
magnetic flux variations, as characteristic of thick MADs
\citep{Tchekhovskoy2011}, are also present at the disc thickness considered here. Assuming that the thinner the disc
the easier it is to saturate the BH with magnetic flux
\citep{Tchekhovskoy2014}, it is likely that our disc is close to the
saturation of the MAD state, since the dimensionless magnetic flux
($\Phi_{\rm BH}/(\dot Mr_{g}^2c)^{1/2}\approx25$) is only a factor of $2$ smaller than the saturation value for thick discs
\citep{Tchekhovskoy2011,Mckinney2012}.

\begin{figure*}
\centering
\includegraphics[width=7.0in,trim=0cm 0cm 0cm 0cm,clip]{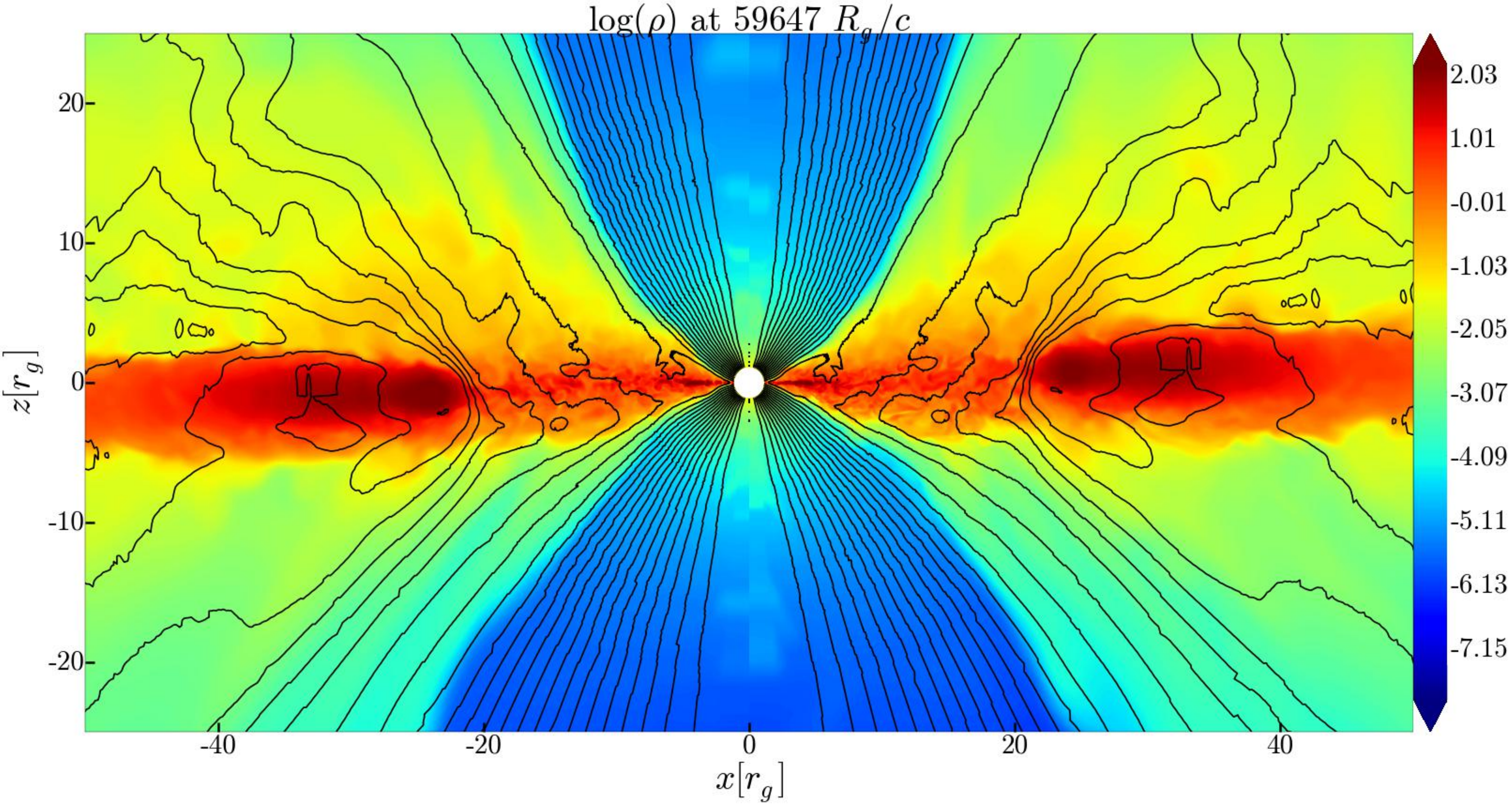}
\caption{Late-time vertical slice through density in our GRMHD simulation of a
  thin disc (compare to Fig.~\ref{fig:bp} at earlier times). The disc
  reaches a steady state where poloidal magnetic flux (black lines)
  gets trapped in the inner disc. The density maximum in the disc at
  $\left|x\right|\sim 25r_g$ marks the transition from the inner low-density,
  high-magnetization disc to the high-density, low-magnetization outer
  disc (see also Fig.~\ref{fig:radplot}a). In Sec ~\ref{sec:disc
    evapor} we argue that the high effective viscosity
  $\alpha_{\rm eff} \sim 1$ in the inner disc may cause radiative cooling to become inefficient and the flow to transition into an ADAF. }
\label{fig:contourplot3}
\end{figure*}

So far we have considered mechanical outflows. To compute the full efficiency, we also need to account for radiative losses in the disc. A rough
estimate for the radiative efficiency is obtained by integrating the
cooling rate, $-\dot{U}$, used to keep our disc thin (at the equilibrium
thickness), from the photon orbit at $r_{\rm photon}=1.43r_g$ to a
large enough radius beyond which the disc luminosity is negligible, e.g.,
$r_{\rm max}=100r_g$, over volume:
\begin{equation}
\eta_{\rm rad}=\frac{-\iiint_{r_{\rm photon}}^{{\rm r_{max}}}
  \dot{U}dV}{\langle \dot{M} c^2 \rangle_{t}},
\label{eq:etarad}
\end{equation}
where $dV = \sqrt{-g}dr d\tilde\theta d\tilde\phi$ is the volume element.  This
gives a radiative efficiency consistent with the \citet{Novikov1973}
value of $\sim 18\%$ (Fig.~\ref{fig:timeplot}(b)), which is encouraging for the
continuum fitting method used to determine BH spin (see e.g.{}
\citealt{Mcclintock2013}). However, as also seen in
Fig.~\ref{fig:timeplot}(b), the total efficiency $\eta_{\rm tot}=\eta_{\rm jet}+\eta_{\rm wind}+\eta_{\rm rad}$  of our disc reaches $60{-}80$\%. 

\subsection{Jet geometry}
\label{sec:jet-geometry}
A major surprise of this work is the finding of powerful jets, even in
our thin disc accretion system. How do the jets from thin discs compare
to the more familiar jets from thick discs? To carry out the
comparison, we estimate the cross-sectional area $A$ of the jet, defined by $p_b>1.5\rho c^2$, at each radius r:
\begin{equation}
A=\iint \Heavi(p_b-1.5\rho c^2) dA_{\tilde \theta \tilde \phi},
\end{equation}
Assuming the jet cross-section is circular, we obtain its
effective half-opening angle as
\begin{equation}
\Delta\theta=\sqrt{\frac{A}{\pi r^2}}.
\end{equation}
Fig.~\ref{fig:tiltvsr}(d) compares the radial dependence of our jet
opening angle to that of a jet from a thick $H/R \sim 0.3$ disc, as found in non-radiative GRMHD
simulations described in \citet{Liska2018A}. The jet opening angle in
the present work is much wider, presumably due to lack of the support in
collimation pressure from the much thinner and cooler ambient medium consisting out of the disc and corona. This wider
opening angle associated with thin discs may lead to differences in
the radiative flux (\citealt{Fragile2012}), possibly explaining the absence of any
observations of soft state jets in XRBs.
Figure~\ref{fig:tiltvst}(c) shows that $\Delta\theta$ decreases over
time, possibly due to a decrease in jet power, as implied by the
decreasing mass accretion rate and approximately constant jet
efficiency, as seen in Fig.~\ref{fig:timeplot}(a),(b). If the power of
the jet is lower, its pressure is also lower, so the ambient medium
would compress the jet into a smaller opening angle.

\subsection{Radial structure}
\label{sec:radialstructure}

\begin{figure}
\centering
\includegraphics[width=\linewidth,trim=0cm 0cm 0cm 0cm,clip]{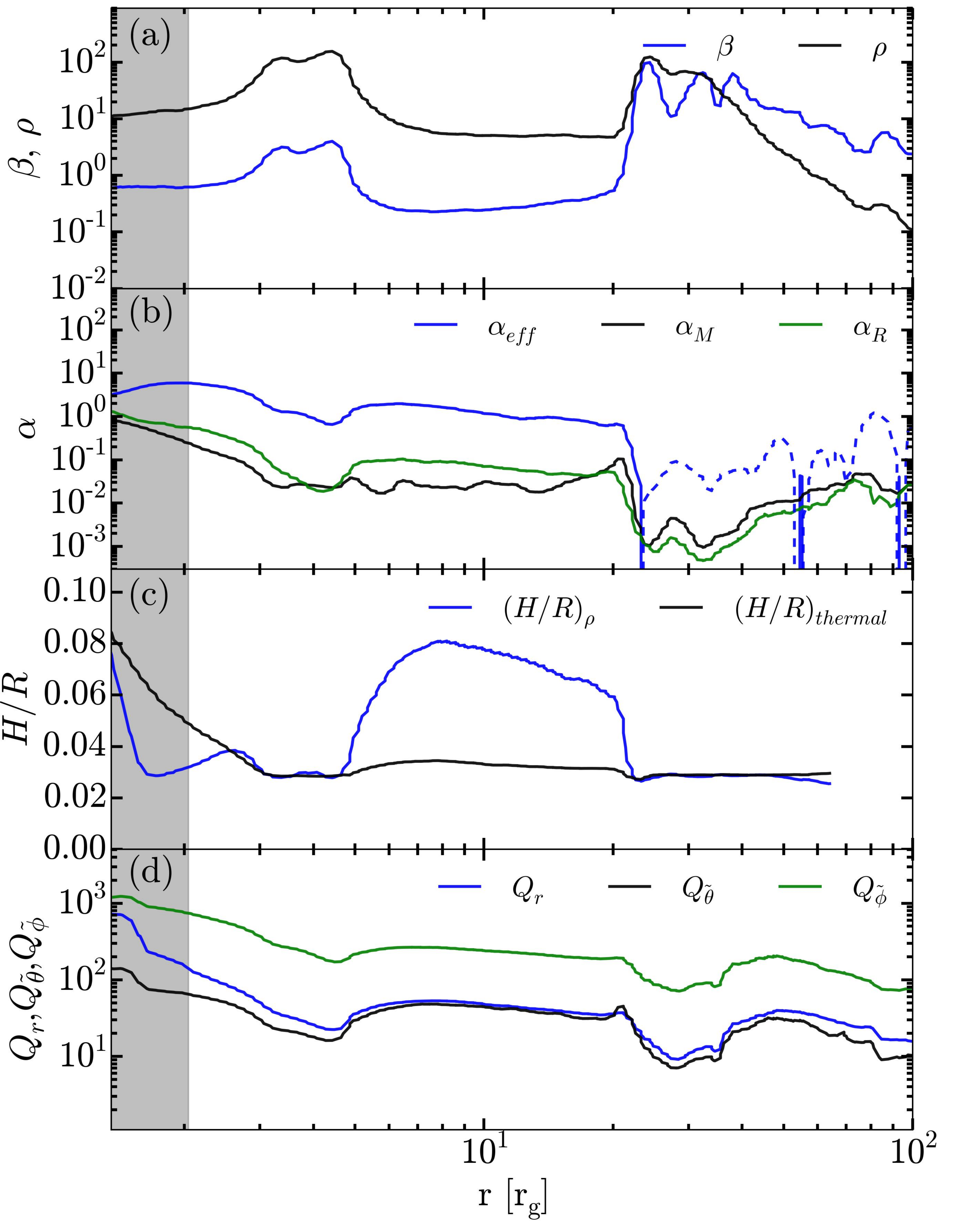}
\vspace{-20pt}
\caption{Radial profiles averaged over
  $6\times10^4<t<6.25\times10^4r_g/c$. Radii within the ISCO ($r_{\rm
    ISCO}\approx2.04 r_g$) are shaded grey. 
  (a) The accretion disc transitions from an inner low-$\beta$ low-density flow to
  an outer high-$\beta$ high-density flow at $r \approx 25r_g$. (b) The radial dependence of the viscous stresses. Since the
  effective viscosity exceeds the Reynolds and Maxwell stresses by one
  order of magnitude, $\alpha_{\rm eff}\gg\alpha_{M,R}$, it is likely
  that large scale torques transport angular momentum outwards. The dotted lines denote negative values of
  $\alpha_{\rm eff}$. The disc is outflowing
  beyond the stagnation surface at $r \approx 23r_g$. (c) Within $25 r_g$ the
  density scale height $(H/R)_{\rho}$ increases above the thermal scale height $(H/R)_{\rm thermal}$ due to
  magnetic pressure support in the inner disc. (d) We maintain more than $10$ cells per MRI wavelength, $Q$, in all
  $3$ dimensions, $r,\tilde \theta,\tilde \phi$ over most of the disc. }
\label{fig:radplot}
\end{figure}

Figure~\ref{fig:contourplot3} shows a vertical slice through a
late-time state of the system. The accretion disc exhibits a sharp
transition in density around $|x|\sim 25r_g$: at smaller radii, the
accretion disc is lower-density and contains a larger magnetic flux
(larger number of magnetic field lines) than at larger radii.
We can see the same more quantitatively in Fig.~\ref{fig:radplot}, which shows disc radial profiles at late time.
The disc reaches a
quasi-steady state with a low-density, low plasma-$\beta$ inner disc
truncated at $r\sim25r_g$ coupled to a high density, high plasma-$\beta$ %
outer disc at larger radii. Here $\beta$ is calculated by taking the ratio of the density weighted average gas pressure $p_g$ and the density weighted average magnetic pressure $p_b$. Note the presence of a
density bump before the ISCO due to mass accumulation, as is
characteristic for thin discs (\citealt{Penna2010}). To
understand how this density and magnetic structure affects the disc
dynamics, we compute the Maxwell viscosity, $\alpha_M$, Reynolds
viscosity, $\alpha_R$, and effective viscosity, $\alpha_{\rm eff}$,
parameters as follows,
\begin{align}
\alpha_M&=\frac{\langle b_{r}b_{\tilde{\phi}}\rangle_{\rho}}{\langle p_{\rm b}+p_{\rm g}\rangle_{\rho}},\\ 
\alpha_R&=\frac{\langle (\rho c^2+u_{\rm g}+p_{\rm g})\delta u_{r} \delta u_{\tilde{\phi}}\rangle_{\rho}}{\langle p_{\rm b}+p_{\rm g}\rangle_{\rho}}\\
  \alpha_{\rm eff}&=\frac{\langle u_{\tilde{\phi}}u_{r}\rangle_{\rho}}{\langle c_{i}^{2}\rangle_{\rho}},\\
  c_{i}&=\sqrt{\frac{p_{\rm g}}{\rho c^2}}.
\end{align}
Here $c_i$ is the isothermal sound speed and $\delta u=u-\langle u \rangle_{\rho}$ is the velocity deviation from the mean. The density-weighted average of a given quantity $\rm x$ is denoted by $\langle \rm x \rangle_{\rho}$. The averaging is performed within $\sim10$ scaleheights of the disc's midplane, more specifically for $\rho>0.01$.

Figure~\ref{fig:radplot}(b) shows that the Reynolds stress exceeds the
Maxwell stress by a factor of $\sim2$ in the inner disc while in the outer disc the Maxwell stress dominates.  
The effective viscosity exceeds the sum of Reynolds and Maxwell viscosity
contribution by an order of magnitude and becomes negative in the corona. This might be indicative of large-scale magnetic
torques/winds removing the angular momentum from the inner disc and is not surprising since we define the effective viscosity based on a 1D stationary $\alpha$-disc model that does not take into account any time-variability and magnetic effects. In fact, deviations
of (GR)MHD results from the $\alpha$-viscosity description are rather
typical
\citep{Avara2016,Texeira2014,Mckinney2012,Penna2010,Sorathia2010} and
indicative of the intrinsic limitations of the $\alpha$-disc
\citep{ss73}, especially at greater field strengths.  It is an
interesting question whether the reported discrepancy \citep{King2007} between the
values for the $\alpha$-viscosity found in local MHD simulations (derived from Maxwell and Reynolds stresses) and
those constrained from observations (derived from the effective viscosity set by
accretion timescale) might be due to the
limitations of the $\alpha$-disc description rather than a mismatch
between the simulations and observations. 

Interestingly, while the Maxwell and Reynolds stresses remain positive throughout the disc, the effective viscosity (in other words $u_r$) becomes negative for $r>23 r_g$, causing the disc to spread out (Fig.~\ref{fig:radplot}(b)). Viscous spreading is characteristic for all finite-size accretion discs and is caused by an outwards flux of conserved angular momentum as mass moves inwards. However, in contrast to the thick $H/R\sim0.3$ discs considered in our previous work \citep{Liska2018A}, where viscous spreading caused the precession to stall, it is unlikely to significantly affect the internal disc dynamics in this work since the timescale for viscous spreading is more than an order of magnitude longer than the simulation runtime.

Due to the global alignment of the disc over time, the tilt angle
evolves from the initial $10^\circ$ to $\sim 2^\circ$ at late
times, and therefore our simulation sweeps through a wide range of tilt angles.
Throughout the evolution, we do not see any evidence of
significant tilt-related effects on the internal disc dynamics, such
as standing shocks/sharp entropy gradients aligned with the lines of
nodes \citep{Fragile2008}. 

We define the thermal and density disc scale heights, respectively, as
\begin{align}
(H/R)_{\rm thermal}&=\frac{\langle c_i \rangle_{\rho^2}}{\langle v_{\tilde{\phi}} \rangle_{\rho^2}},\\
(H/R)_{\rm \rho}&=\langle|\tilde{\theta}-\tilde{\theta}_{avg}|\rangle_{\rho}
\end{align}
Here $c_i$ is the isothermal sound speed and $v$ is 3-velocity, and
$\tilde{\theta}_{avg}$ is the average $\theta$-position of the disc's
midplane in tilted coordinates. Figure~\ref{fig:radplot}(c) shows that
while the thermal scale height remains approximately constant for $r>3 r_g$, $(H/R)_{\rm thermal}\approx 0.03$, the density scale
height develops a bump, $(H/R)_\rho\sim0.08$ at $r\sim 10r_g$,
due to the excess magnetic pressure support in a strongly magnetized,
$\beta\ll1$, disc. Around the ISCO the scaleheight increases, because the disc becomes super Keplerian, while the cooling function assumes a Keplerian disc.

To quantify the degree to which our simulation resolves the magnetized
turbulence in the disc, we compute the quality factors
$Q_{r,\tilde \theta,\tilde \phi}$, which give the number of cells per MRI wavelength
in each of the three directions: 
\begin{align}
\label{eq:Qmri}
Q_{r,\tilde \theta,\tilde \phi}&=\frac{2\pi}{\Delta_{r,\tilde \theta,\tilde \phi}}\frac{\langle
                   v^{\rm A}_{r,\tilde \theta,\tilde \phi} \rangle_w}{\langle \Omega \rangle_w},\\
v^{A}_{r,\tilde\theta,\tilde\phi}&=\sqrt{\frac{b_{r,\tilde{\theta},\tilde{\phi}}^2}{(\rho c^2+u_{\rm g}+p_{\rm g}+b^2)}},\\
\Omega&=\frac{v_{\tilde{\phi}}}{r},
\end{align}
where $\Delta_{r,\tilde \theta,\tilde \phi}$ is the size of the cell.  The Alfven
speed, $v_{\rm A}$, and angular frequency, $\Omega$, are
volume-averaged with $w=\sqrt{b^2\rho}$ as the weight since a large fraction
of the mass in our thin discs resides in equatorial current sheets
where the magnetic field vanishes.  Figure~\ref{fig:radplot}(d) shows
that the whole disc is well resolved with $Q_{r,\tilde \theta}>30$, $Q_{\tilde \phi} >
200$ in the inner disc and
$Q_{r,\tilde \theta}>10$, $Q_{\tilde \phi}>70$ over most of the outer
disc. This satisfies the numerical convergence criteria for MRI turbulence, $Q_{\theta}>10$ and $Q_{\phi}>20$ (see
e.g. \citealt{Hawley2011}). 
\section{Discussion}
\label{sec:Discussion}
\subsection{Bardeen-Petterson Alignment}
\label{sec:disc-bp}
We found that the inner $\sim5r_g$ of a
thin disc, $H/R\simeq 0.03$, initially tilted by $10^\circ$ relative to the
central spinning BH, undergoes alignment
with the BH equator (Sec.~\ref{sec:bp}). This is the first demonstration of the
\citet{bp75} effect in a GRMHD simulation, in the presence of non-local and anisotropic turbulent MHD
stresses.  

This confirmation of the BP effect has profound consequences for the growth and spin evolution of supermassive BHs (SMBHs), since BP alignment is a crucial ingredient that has been \textit{assumed} to take place for misaligned accretion episodes (e.g. \citealt{Volonteri2005,King2006,Fanidakis2011}). 
Because the alignment radius acts as lever arm helping to torque the
BH, the BP effect can torque the BH and align its spin vector with the outer misaligned
accretion flow on much shorter timescales than otherwise
\citep{Rees1978, Scheuer1996, Natarajan1998}. This rapid reorientation
of BH spin has the potential to create the right conditions for rapid
BH spin-up.
If BH spin reorientation occurs on a shorter timescale than the
timescale of a single accretion episode in a
chaotic accretion scenario (in which the direction of the supplied gas
angular momentum randomly changes between different accretion
episodes, see e.g.{} \citealt{Volonteri2005}), then supermassive BHs
can be efficiently spun up (e.g.
\citealt{Natarajan1998}). In the opposite case, the
accretion-supplied angular momenta would tend to cancel out, and the
central BHs would be on average spun down.

Due to the high cost of GRMHD simulations, it is appealing to use them to calibrate both computationally cheaper SPH simulations and analytic theory. Our simulations are in the viscosity-dominated,  $H/R<\alpha$
regime \citep{Papaloizou1983} since both our effective viscosity parameter,
$\alpha_{\rm eff} \simeq 1$, and local Maxwell plus Reynolds stress
related viscosity parameter, 
$\alpha_M+\alpha_R \lesssim 0.1$, exceed $H/R=0.03$ (see
Figure~\ref{fig:radplot}). However, we did not find Bardeen-Petterson
alignment in any of our simulations featuring thicker discs with $<H/R=0.1$ (\citealt{Liska2019A}). Because all of these simulations have
$\alpha_{M,R} \lesssim H/R< \alpha_{\rm eff}$, it appears likely that values
of local Maxwell and Reynolds stresses determine the transition from
the wave- ($H/R>\alpha$) to the viscosity- ($H/R<\alpha$) dominated
regime, where BP alignment is expected.

However, the alignment radius we find substantially differs from both the
analytical theory and SPH simulations. Specifically, \citet{Kumar1985}
found analytically using the corrected \citet{bp75} equations in
\citet{Papaloizou1983} a radius $r_{\rm BP} \approx 300$ $r_g$, while
\citet{Nelson2000} find $r_{\rm BP} \approx 100$ $r_g$ for $\alpha=0.05$
and $h/r=0.03$. SPH simulations of \citet{Nelson2000} find
$r_{\rm BP} \approx 30r_g$ for $\alpha=0.1$ and $h/r=0.03$. Other more
recent SPH work considers a smaller disc thickness $h/r\approx0.013$
and smaller viscosity $\alpha\approx0.03$ \citep{Lodato2010}, which makes
comparison to our work difficult.  However, assuming a
scaling relation, $r_{\rm BP}\sim [(h/r)^{-2}/\alpha]^{4/7}$
\citep{Kumar1985}, these results are in similar
disagreement.

One contributing factor to this disagreement with SPH simulations may
be the $\sim 2$ times larger density scale height, caused by the buildup of
magnetic pressure, between $5r_g<r<20r_g$ in our simulations (see
Sec.~\ref{sec:radialstructure}). We indeed see that $r_{\rm BP}$ (Fig.~\ref{fig:tiltvsr}a) coincides with the radius where the disc becomes thicker (Fig.~\ref{fig:radplot}(c)). Naively, analytically one would only predict \citep{Kumar1985} an
$\approx 0.4$ times smaller alignment radius for a similar increase in disc thickness, insufficient to account
for the full extent of the discrepancy. However since $H/R \approx \alpha_{R}+\alpha_{M}$ the disc may not be in the $H/R<\alpha$ viscosity dominated regime required for BP alignment. Another contributing factor
might be the presence of large-scale magnetic torques in the system,
which can affect the alignment radius in at least two ways. First, these torques might induce coupling between the inner
aligned disc and outer misaligned disc-corona-jet system moving $r_{\rm BP}$ inwards. Second, as
discussed in Sec.~\ref{sec:radialstructure}, large-scale magnetic
torques can remove angular momentum from the disc and increase the
inflow velocity. Because BP alignment is expected to occur more
rapidly when the radial inflow velocity is smaller \citep{bp75}, this
might reduce the efficiency of BP effect (\citealt{Nealon2015}). 

To test if an external torque can explain the smaller-than-predicted BP alignment radius, we calculate the ratio between the external magnetic torque dragging misaligned angular momentum inwards and the LT torque. Note that since $\alpha_M+\alpha_R\ll\alpha_{eff}$  (Fig.~\ref{fig:tiltvsr}(c)) we can safely neglect internal stresses and assume that accretion is driven solely by an external magnetic torque. This viscous torque is given by (e.g. \citealt{ss73}),
\begin{equation}
T_{\rm mag}=\alpha_{\rm eff}(h/r)^2 \times v_{\rm K} \times T^r_{\tilde{\phi}} \times \sin(\mathcal{T}),
\end{equation}
while the LT torque is given by, 
\begin{equation}
T_{\rm LT}=T^r_{\tilde{\phi}} \times \sin(\mathcal{T}) \times \Omega_{\rm LT}, 
\end{equation}
 Assuming $\alpha_{eff} \sim 1$ and using eq.~\ref{eq:prec} we conclude that these two torques are equal around $r \sim 15 r_g$. This still overestimates $r_{\rm BP}$ by a factor $\sim 3$, but suggests that large scale torques may indeed contribute to the discrepancy between our work and SPH simulations. Note that in this very crude calculation we neglected that the LT torque acts perpendicular to the magnetic torque. Namely, $T_{\rm LT} \gtrsim T_{\rm mag}$ does not guarantee BP alignment, since the disc may keep precessing as a rigid body without aligning (Liska et al 2019b, in prep). For BP alignment, misaligned angular momentum also needs to mix azimuthally such that net alignment is produced \citep{Sorathia2013}. This mixing may take place on timescales (much) longer than the viscous time and thus explain this remaining factor $\sim 3$ discrepancy.

 A smaller $r_{\rm BP}$ could have a significant effect on the predictions of SMBH growth models because smaller values of $r_{\rm BP}$ lead to less rapid alignment between BH and outer disc and perhaps consequently less rapid spin up. For the same reason, our result implies that initially misaligned X-ray binary systems will take even longer to align than previously predicted \citep{King2016}, indicating that there could be many misaligned X-ray binaries today, as implied by the Lense-Thirring precession QPO model of \citet{Ingram2009}.
Future work should study the effect of BP alignment and disc warp on the
measured values of BH spin \citep[e.g.,][]{Mcclintock2013}. 

The inclusion of a gas pressure dominated equation of state with
adiabatic index $\Gamma=5/3$ as in this work is only applicable in the
outer accretion disc of X-ray binaries (see e.g.{}~\citealt{zhu2013}),
while the inner part could be radiation pressure dominated with
$\Gamma=4/3$. We find that changing the adiabatic index to
$\Gamma=4/3$ for a thin disc tilted by $45^\circ$ behaves qualitatively
similar to $\Gamma=5/3$ (Liska et al 2019B, in prep). This is not unexpected since the most
dominant effect of a softer equation of state is that the disc becomes
thinner for a given specific internal energy $u_g/(\rho c^2)$. However, the cooling function \citep{Noble2009}
automatically adapts to the equation of state in order to maintain
the desired thermal scale height and thus the absence of any strong
dependence on the adiabatic indices is not unexpected. 

\subsection{Disc Evaporation?}
\label{sec:disc evapor}
The transition from a low-viscosity, high-density outer disc into a
high-viscosity, low-density inner disc (See
Sec.~\ref{sec:radialstructure} and Fig.~\ref{fig:contourplot3}/\ref{fig:radplot}) might provide clues into a
long-standing puzzle in accretion physics: How do cool thin discs \citep{ss73}
transition into hot thick radiatively-inefficient accretion flows \citep{Narayan1994} near the black hole? This work shows that magnetically driven winds can lower the disc density and increase the inflow speed with respect to the outer disc. Subsequently, the ions and electrons may become weakly coupled and the cooling timescale may become limited by the timescale for Coulomb collisions and other plasma processes to equilibrate the temperature of the hot non-radiative ions with the radiatively-cooled synchrotron emitting
electrons \citep{Shapiro1976}. This may prevent the inner disc from cooling and can conceivably lead to a
radiatively inefficient thick accretion flow at a radius $r\lesssim 25r_g$ for this setup. Indeed, the finite timescale for electron-ion coupling implies that one would generally expect such a disc to form for $\dot{M}<\alpha^2\dot{M}_{\rm Edd}$ (e.g. \citealt{Esin1997}). This disc-evaporation mechanism, through the elevated $\alpha$-viscosity in the inner disc, is attractive in that it does not require conduction of heat from the corona to the disc (e.g. \citealt{Meier1994,Liu1999,Czerny2000, Qian2007}).

How does the high-viscosity, low-$\beta$ inner disc, seen in
Figure~\ref{fig:radplot}(a), form?  In our simulation it may have formed
due to the rapid cooling of an initial torus threaded with poloidal
magnetic flux \citep{Sikora2013,Begelman2014}: the cooling causes the thermal
pressure to decrease, but -- due to vertical magnetic flux
conservation -- the magnetic flux stays about the same. This causes
the disc to become more strongly magnetized and plasma $\beta$ to
drop. 
Shearing box simulations seeded with strong vertical magnetic flux
appear to develop a similarly highly magnetized accretion state with
strong outflows \citep{Salvesen2016,Bai2012}.  If this
scenario is indeed the case, it would require the presence of large
scale magnetic flux in the accretion disc prior to the disc becoming
thin, which would limit the applicability of this simulation to the
intermediate states for X-ray binaries.

Future work will investigate the effect of the different initial
magnetic field geometries \citep[e.g.,][]{2018arXiv180904608L}, exploring if large scale poloidal magnetic flux is indeed a necessary ingredient for the high $\alpha$-viscosity inner disc.
It will also include electron-ion coupling, and on-the-fly radiation transfer,
to accurately model the cooling of the disc.

\subsection{Jet Launching}
\label{sec:jetlaunch}
This work shows that thin discs down to at least $H/R=0.03$ can
efficiently launch relativistic \citet{bz77} jets of substantial
power, carrying out $\gtrsim20\%$ of the accretion power, over 
timescales comparable to the accretion
time (Sec.~\ref{sec:jet-geometry}). %
This suggests that even such thin discs as considered in this work are
capable of retaining for their accretion time large-scale poloidal (vertical) magnetic flux on
the BH, a necessary ingredient for launching relativistic jets
\citep{bz77}. This is particularly interesting given that simple
analytical arguments suggest that thin discs should lose their
large-scale magnetic flux to outward diffusion \citep{Lubow1994}.
It is possible that the large
scale external torques may overcome this problem by
dragging flux inwards before it has time to diffuse out (see also
\citealt{Giulet2012,Giulet2013}).  

How can we reconcile the formation of powerful jets from thin discs
with observations? There are no observations that have
convincingly detected jets from thin discs in the soft state of X-ray
binaries (though see \citealt{Rushton2011}), however, about $10\%$ of
quasars are radio loud and form jets \citep{Sikora2007}. Because our simulated jets have wide opening angles,
$\Delta\theta\sim20^\circ$ (Fig.~\ref{fig:tiltvsr}c), they might
become less optically thick and more difficult to detect (see
also \citealt{Russel2011, Fragile2012}). Another possible explanation is that our simulations do not apply to the soft state of XRBs (Sec.~\ref{sec:transdisc}).

\subsection{A Transitional Disc?}
\label{sec:transdisc}
An interesting possibility is that our simulations apply to \emph{transitional} discs, in the middle of the hard-to-soft state transition (e.g. \citealt{Fender2004}). In fact, we set up our simulations in a very similar way: the initial thick torus rapidly cools down to the target thickness, $H/R=0.03$, which is much smaller than the initial thickness, $H/R\sim 0.3$. Since thick accretion discs may be able to generate and advect large scale poloidal magnetic flux through large scale dynamo action \citep[e.g.,][]{2018arXiv180904608L}, they are expected to retain a substantial amount of it after their collapse into a thin disc. This can lead both to a highly viscous inner disc that evaporates into an ADAF (Sec.~\ref{sec:disc evapor}) and sustains a strong jet (Sec. \ref{sec:jetlaunch}, see also \citealt{Ferreira1993, Fereira2006,Sikora2013,Begelman2014}). Indeed XRBs in the hard-to-soft state transition are known to produce jets (e.g. \citealt{Fender2004}), while radio-loud quasars may contain such transitional discs \citep{2015ASSL..414...45T}. Spectral modeling of two-temperature magnetically truncated discs has proven successful in explaining both emission in X-Ray and radio during XRB state transitions \citep{Marcel2018A, Marcel2018B}.

However, as proposed in \citet{Lubow1994} this flux may slowly diffuse out and cause the jet to shut down. In addition, if this large scale poloidal magnetic field indeed leads to evaporation of the inner disc into an ADAF (Sec.~\ref{sec:disc evapor}), the truncation radius between the inner thick and outer thin disc will move inwards. This is consistent with observational evidence of the truncation radius moving in during the evolution towards the soft state \citep{Esin1997,Done2007,Ingram2011}. 
Figure~\ref{fig:timeplot}(c) indeed shows signs of magnetic flux diffusing out of the BH: the flux in the disc ($\Phi_{\rm disc}$) stays roughly constant while the flux on the BH ($\Phi_{\rm BH}$) drops. However, the drop is small and appears to be leveling off. 
Several mechanisms have been suggested that can prevent the poloidal magnetic flux from diffusing out in thin discs \citep{Rothstein2008,Giulet2012,Giulet2013}. Future simulations spanning much longer runtimes can probe if thin discs are able to retain poloidal magnetic flux for a more extended time period, or are always transitional.

\section{Conclusions}
\label{sec:conclusions}
In this work we have performed the thinnest disc GRMHD simulations to
date. We started with an $H/R=0.03$ accretion disc tilted by
$10^\circ$ relative to a rapidly spinning $a=0.9375$ BH. Using 3 AMR
levels, we carried out GRMHD simulations at sufficiently high effective
resolution, $2880\times864\times1200$, which for the first time
resolved the MRI turbulence in a thin disc in all 3 dimensions with
near-cubical cells (of order unity aspect ratio). Our results can be
summarized in 3 key points.

First, we have confirmed for the first time that the inner parts of
tilted thin discs can align with the BH equatorial plane as theorized
40 years ago by \citet{bp75}, even when the full effects of GR,
anisotropic MRI turbulence and torquing of the disc by magnetized
corona and jets are included. The disc aligns with the BH within the
BP radius, $r_{\rm BP}\simeq 5r_g$, whose value is expected to increase for thinner discs (e.g. \citealt{Kumar1985}). The development of a BP
configuration can have profound consequences for the evolution of BH spins
in AGN, as the large lever arm of $r_{\rm BP}$ out to which the disc
is aligned can torque the BH into alignment with the outer, tilted
disc on a much shorter timescale than without the BP effect (e.g. \citealt{Scheuer1996,Natarajan1998}). 

Second, we have shown that an accretion disc can develop an inner
low-density, high-viscosity disc coupled to an outer high-density,
low-viscosity disc at $r\lesssim 25r_g$. We suggested that the order unity viscosity of
the inner disc we find might lead to it evaporating into a radiatively
inefficient accretion flow when the electron-ion coupling time exceeds the accretion time (e.g. \citealt{Esin1997}). This high viscosity may be caused by the presence in the initial conditions of large-scale poloidal magnetic flux, which removes the angular momentum through large-scale outflows. Large scale poloidal magnetic flux may be present in thin discs during hard-to-soft state transitions (e.g. \citealt{Sikora2013,Begelman2014}), as discussed in Sec.~\ref{sec:transdisc}.

Third, we have shown that BH accretion systems with thin discs, if
initially threaded with large scale poloidal magnetic flux, can launch powerful
\citet{bz77} jets on the viscous timescale, with their power reaching
$20{-}50\%$ of the accretion power. This challenges the standard
paradigm that thin discs in the soft state cannot advect inwards poloidal
magnetic flux needed to launch jets \citep{Lubow1994} and is seemingly in tension with
the lack of any clear detection of jets in X-ray binaries. However
the morphology of our jets, specifically their twice as large opening
angle as of those produced by thick discs
(e.g. \citealt{Liska2018A,Mckinney2006}; Chatterjee et al 2018a, in
prep), may make them more optically thin and thus more
difficult to detect (see also \citealt{Fragile2012}).
Another possibility is that our simulations describe transitional discs in the hard-to-soft state transition which are known to produce powerful jets (e.g. \citealt{Fender2004}) and, like our simulations, may naturally harbor large scale poloidal magnetic flux \citep{Sikora2013}, which is required to produce powerful jets \citep{bz77}. This flux may eventually diffuse out (e.g. \citealt{Begelman2014}) causing the jets to shut down. Outwards flux diffusion might indeed be present in our simulation  (Sec.~\ref{sec:transdisc}).

\section{Acknowledgments}
\label{sec:acks}
We thank Chris Fragile and Cole Miller for useful suggestions. AI thanks James Matthews for useful discussions. This research was made possible by NSF PRAC award no.~1615281 and OAC-1811605 at the
Blue Waters sustained-petascale computing project and supported in
part under grant no.~NSF~PHY-1125915. ML and MK were supported by the
Netherlands Organisation for Scientific Research (NWO) Spinoza Prize, AI by the Royal
Society URF, AT by Northwestern University and the TAC and NASA Einstein (grant no.~PF3-140131)
postdoctoral fellowships.

\section{Supporting Information}
Additional Supporting Information may be found in the online version
of this article: movie file (\href{https://www.youtube.com/watch?v=eccVsm04_SM}{link}).

\bibliography{mybib,sasha}
\bibliographystyle{mn2e}
\label{lastpage}
\end{document}